\documentstyle[sprocl,epsf]{article} 

\newcommand\gae{\stackrel{>}{\sim}}

\newcommand\iic{\ \ ,}
\newcommand\iip{\ \ .}

\newcommand\gev{{\rm GeV}}
\newcommand\be{\begin{equation}}
\newcommand\ee{\end{equation}}
\newcommand\bea{\begin{eqnarray}}
\newcommand\eea{\end{eqnarray}}

\begin{document} 
 
\title{COLORONS: THEORY AND PHENOMENOLOGY\footnote{Talk given on November
13, 1996 at the 1996 International Workshop on Perspectives of Strong
Coupling Gauge Theories (SCGT96), Nagoya, Japan.}} 
 
\author{E.H. SIMMONS } 
 
\address{Department of Physics, Boston University \\
590 Commonwealth Avenue, Boston MA 12215, USA}

 
\maketitle

\bigskip
\begin{picture}(0,0)(0,0)
\put(295,250){BUHEP-97-1}
\put(295,235){hep-ph/9701282}
\end{picture}

\abstracts{ We briefly describe the structure and  phenomenology
of a flavor-universal extension of the strong interactions, focusing on the
color-octet of massive gauge bosons (`colorons') present in the low-energy
spectrum.  We discuss current limits on the colorons and what future
measurements may reveal.}


\section{Introduction}

Data from the LEP and SLD experiments\cite{WARSAW96} is in such good agreement with the
standard model that it tightly constrains non-standard electroweak physics.  Yet
the existence of significant non-standard strong interactions remains possible. 
Indeed an apparent excess of high-$E_T$ jets has been found in the inclusive jet
spectrum measured by CDF \cite{CDFexc}.  Finding a model that is consistent with
both the LEP and Tevatron findings is an interesting challenge.

In this context, a flavor-universal coloron model \cite{newint} has been 
proposed.  This model is a flavor-universal variant of the coloron model of Hill
and Parke\cite{topglu} which accommodates the jet excess without
contradicting other experimental data.  It involves a minimal
extension of the standard description of the strong interactions,
including the addition of one gauge interaction and a scalar
multiplet, but no new fermions.  As such, it serves as a useful
baseline with which to compare both the data and other candidate
explanations of the jet excess \cite{pdfs}.  

We will briefly describe the structure and phenomenology of the Higgs phase
of the model (for a fuller discussion see ref. \cite{colph}).  We
discuss current limits on the colorons and what future measurements may reveal.

\section{The model}
\label{sec:model}

In the flavor-universal coloron model \cite{newint}, the strong gauge
group is extended to $SU(3)_1 \times SU(3)_2$.  The gauge couplings
are, respectively, $\xi_1$ and $\xi_2$ with $\xi_1 \ll \xi_2$.  Each
quark transforms as a (1,3) under this extended strong gauge group.
The model also includes a scalar boson $\Phi$ transforming as a
$(3,\bar 3)$.  In the model's Higgs phase, the
scalar develops a non-zero vacuum expectation value $\langle\Phi\rangle$ at a
scale where neither gauge coupling is strong.  This vev breaks the two strong
groups to their diagonal subgroup, which is identified with QCD.

The original gauge bosons mix to form an octet of massless gluons and
an octet of massive colorons.  The gluons interact with quarks through
a conventional QCD coupling with strength $g_3$.  The colorons
$(C^{\mu a})$  have a vectorial interaction with quarks 
\be
{\cal L} = - g_3  \cot\theta J^a_\mu C^{\mu a} \iic
\ee
where  $J^a_\mu \equiv \sum_f {\bar q}_f \gamma_\mu
\frac{\lambda^a}{2}q_f$ and $\cot\theta = \xi_2/\xi_1\, $.  Note that we
expect $\cot\theta >1$.  In terms of the QCD coupling, the gauge boson mixing
angle and the scalar vev, the mass of the colorons is
\be
M_c = \left( \frac{g_3}{\sin\theta \cos\theta} \right) \langle\Phi\rangle \iip
\ee
The colorons decay to all sufficiently light quarks; assuming
there are 6 flavors lighter than $M_c/2$ and writing $\alpha_s \equiv
g_3^2/4\pi$, the decay width is
\be
\Gamma_c \approx \alpha_s \cot^2\theta\, M_c \iip
\label{eqwid}
\ee

This model has several appealing features.  The extended
strong interactions can be grafted onto the standard one-Higgs-doublet
model of electroweak physics, yielding a simple, complete, and
renormalizable theory.  The flavor-universality of the new strong interactions
prevents the introduction of new flavor-changing neutral currents.  Corrections
to the $\rho$ parameter are small (see section 3).  Finally, the model
fulfills its original purpose of accommodating the jet excess.   Writing the
low-energy interaction among quarks that results from heavy coloron exchange as
a four-fermion interaction
\be
{\cal L}_{4f} = - \frac{g_3^2 \cot^2\theta}{M_c^2} J^a_\mu J^{a \mu}
\label{fourff}
\ee
and including its the contributions to the inclusive jet cross-section (setting
$M_c / \cot\theta = 700$ GeV) 
yields the curve in figure 1, which
compares quite nicely with the data.

\begin{figure}[htb]
\vspace{-3.5cm}
\epsfxsize 8cm \centerline{\epsffile{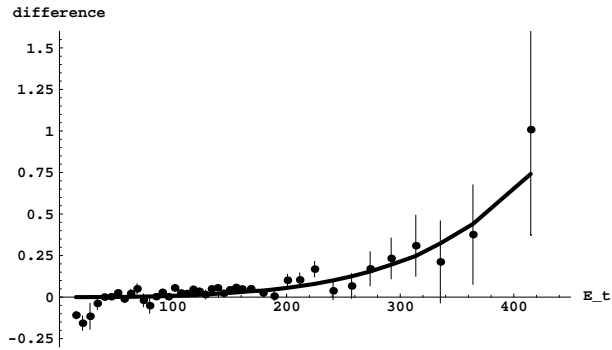}}
\vspace{-3cm}
\caption[differ]{Difference plot ((data - theory)/theory) for the
inclusive jet cross-section ${1\over{\Delta\eta}} \int
(d^2\sigma/d\eta\, dE_T) d\eta $ as a function of transverse jet
energy $E_t$, where the pseudorapidity $\eta$ of the jet falls in the
range $0.1 \leq \vert\eta\vert \leq 0.7$. Dots with (statistical)
error bars are the recently published CDF data \protect\cite{CDFexc}.
The solid curve shows the LO prediction of QCD plus the contact
interaction approximation to coloron exchange of equation
(\protect\ref{fourff}) with $M_C/\cot\theta = 700$ GeV.  Following
CDF, we employed the MRSD0' structure functions \protect\cite{mrs_pak}
and normalized the curves to the data in the region where the
effect of the contact interactions is small (here this region is $45 <
E_T < 95$ GeV).}
\label{diffc}
\end{figure}
%

\section{Existing limits on colorons}
\label{sec:lim-now}

\begin{figure}[htb]
\vspace{-3.cm}
\epsfxsize 8cm \centerline{\epsffile{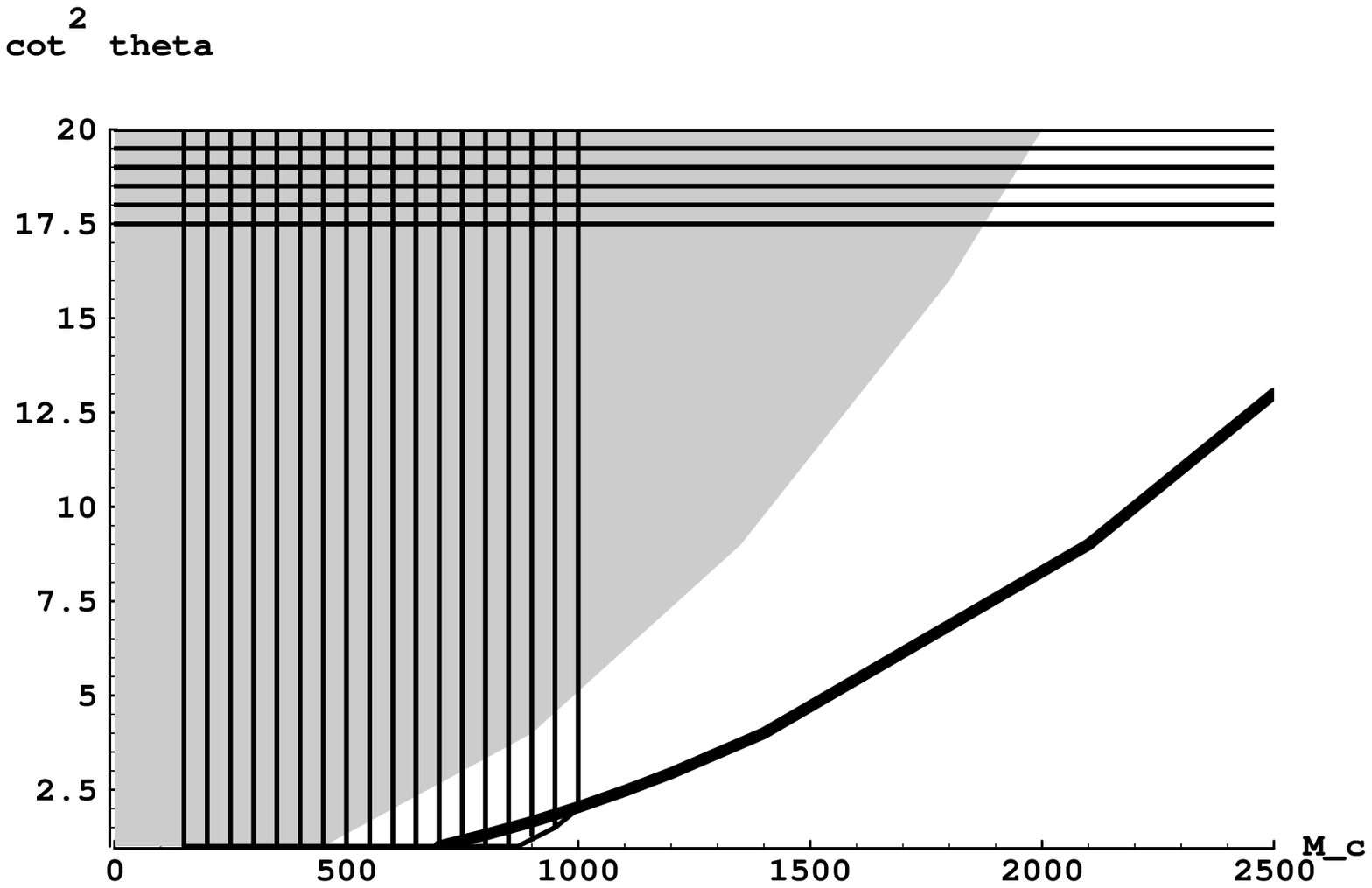}}
\vspace{-3cm}
\caption[allcurlim]{Current limits on the coloron parameter space: mass ($M_c$)
vs. mixing parameter ($\cot^2\theta$).  The shaded region is excluded
by the weak-interaction $\rho$ parameter \cite{newint} as in equation
\ref{rhoweq}.  The vertically-hatched polygon is excluded by searches
for new particles decaying to dijets \cite{CDFdij,ua1}. The
horizontally-hatched region at large $\cot^2\theta$ lies outside the
Higgs phase of the model.  The dark line is the curve $M_c /
\cot\theta = 700$ GeV for reference.} 
\label{allcurlim}
\end{figure}

A sufficiently light coloron would be visible in direct production at
the Tevatron.  The CDF Collaboration has searched for new
particles decaying to dijets\footnote{Because the colorons couple to all 
flavors of quarks, they should also affect the sample of b-tagged dijets
observed at Tevatron experiments. As discussed in ref. \cite{colph}, however,
the limit on colorons from b-tagged dijets will probably be weaker than that
from the full dijet sample.} and reported \cite{CDFdij} a 95\% c.l.
upper limit on the incoherent production of such states.  As
discussed in \cite{colph}, we calculated $\sigma \cdot B$ for colorons, and
found find that for $\cot\theta = 1$, the range $200\ {\rm GeV} < M_c < 870\
{\rm GeV}$ is excluded; at $\cot\theta = 1.5$, the upper limit of the
excluded region rises to roughly 950 GeV; at $\cot\theta \gae 2$, it
rises to roughly 1 TeV.  Realizing that $\sigma \cdot B$ is the same
for a coloron with $\cot\theta = 1$ as for an axigluon \cite{axig} of identical
mass\cite{colph} extends the excluded range of coloron masses.
Axigluons with masses between 150 and 310 GeV are excluded by UA1's analysis
\cite{ua1} of incoherent axigluon production; by extension, colorons in this 
mass range with $\cot\theta \geq 1$ are also excluded.  The combined excluded
ranges of $M_c$ are
\bea
150 \gev < &M_c& < 870 \gev \ \ \ \ \ \ \cot\theta = 1\nonumber \\
150 \gev < &M_c& < 950 \gev \ \ \ \ \ \ \cot\theta = 1.5 \\
150 \gev < &M_c& < 1000 \gev \ \ \ \ \ \cot\theta \gae 2\iic\nonumber
\eea
as summarized by the shaded region of figure 2.

An additional limit on $\cot\theta$ comes from 
constraints on the weak-interaction $\rho$-parameter.
Coloron exchange across virtual quark loops contributes to
$\Delta\rho$ through the isospin-splitting provided by the difference
between the masses of the $t$ and $b$ quarks.  Limits on this type
of correction \cite{cdt} imply that \cite{newint}
\be
\frac{M_c}{\cot\theta} \gae 450 {\rm GeV} \iip
\label{rhoweq}
\ee
This excludes the hatched region of the $\cot^2\theta$ -- $M_c$ plane
shown in figure 2. 

Finally, we mention a theoretical limit on the coloron parameter
space.  While the model assumes $\cot\theta > 1$, the value of
$\cot\theta$ cannot be arbitrarily large if the model is in the
Higgs phase at low energies.  Starting from the low-energy four-fermion
interaction (\ref{fourff}) resulting from heavy coloron exchange, 
we use the NJL approximation to estimate the critical value of
$\cot^2\theta$ as
\be
(\cot^2\theta)_ {crit} = \frac{2\pi}{3 \alpha_s}  \approx 17.5
\ee
This puts an upper limit on $\cot^2\theta$ as indicated in figure 2.

\section{Upcoming limits from Tevatron data}
\label{sec:lim-when}

For colorons weighing a little more than a TeV -- those that are just
above the current dijet mass bound of figure \ref{allcurlim} -- it is 
appropriate to use the cross-sections for full coloron exchange \cite{colph}
when making comparisons with the data.  Such colorons are light enough that 
their inclusion yields differential cross-sections of noticeably different shape
than the four-fermion approximation would give (see figure 3).  Once the
full coloron-exchange cross-sections are employed, the mass and mixing
angle of the coloron may be varied independently.  In particular, one
may study the effects of light colorons with small values of
$\cot^2\theta$, extending the range of accessible parameter
space.

In principle, a detailed analysis (including propagating colorons) of the
inclusive jet spectra measured by CDF and D0 should provide a lower bound on
the coloron mass and coupling.  In practice, however, the limit obtained from the
inclusive jet spectrum will depend strongly on which structure functions are
used to calculate the theoretical cross-sections.  For instance, the new CTEQ
structure functions\cite{pdfs}, which reduce the apparent jet excess, would give
stronger limits on colorons than the MRSD0' structure functions we have
employed.  
 
An analysis that is more independent of structure functions could be based,
instead, on the dijet angular distribution.   Some non-standard strong
interactions would produce dijet angular distributions like that of QCD; others
predict distributions of different shape.  In terms of the angular variable
$\chi$ 
\be
\chi = \frac{1 + \vert\cos\theta^*\vert}{1 - \vert\cos\theta^*\vert}
\ee
QCD-like jet distributions appear rather flat while those which are
more isotropic in $\cos\theta^*$ peak at low $\chi$ (recall that
$\theta^*$ is the angle between the proton and jet directions).  The
ratio $R_\chi$
\be
R_\chi \equiv \frac{N_{events}, 1.0 < \chi < 2.5}{N_{events}, 2.5 <
\chi < 5.0}
\ee
then captures the shape of the distribution for a given sample of
events, e.g. at a particular dijet invariant mass.

The CDF Collaboration has made a preliminary analysis of the dijet
angular information in terms of $R_\chi$ at several values of dijet
invariant mass \cite{rmh}.  The preliminary data appears to be
consistent either with QCD or with QCD plus a color-octet four-fermion
interaction like (\ref{fourff}) for $M_c/\cot\theta = 700$ GeV. Our
calculation of $R_\chi$ including a propagating coloron gives results
consistent with these.  It appears that the measured angular
distribution currently allows the presence of a coloron and will
eventually help
put bounds on $M_c$ and $\cot\theta$.
 
\begin{figure}[htb]
\epsfxsize 6cm \centerline{\epsffile{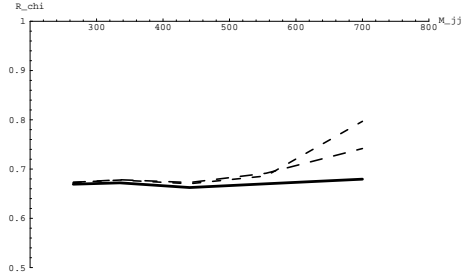}}
\vspace{.2cm}
\caption[shape]{Plot of $R_{\chi}$ as a function of dijet invariant mass showing
  the effects of propagating colorons of different masses when the ratio $M_c /
  \cot\theta$ is fixed at 700 GeV.  The solid curve is the prediction of QCD. The short-dashed
  curve (upper) is for a light coloron: $M_c = 1050$ GeV, $\cot\theta =
  1.5$.  The long-dashed curve corresponds to much heavier colorons
  ($M_c >1400$ GeV) with correspondingly larger values of $\cot\theta$.
  Only the cross-section for the heavier colorons is well-approximated
  by the contact interaction (4) at Tevatron energies.}
\label{shape}
\end{figure}

\section{Conclusions and Prospects}
\label{sec:concl}

The flavor-universal coloron model can accommodate an excess at the
high-$E_t$ end of the inclusive jet spectrum at Tevatron energies
without contradicting other data.  Previous measurements of the
weak-interaction $\rho$ parameter and searches for new particles
decaying to dijets imply that the coloron must have a mass of at least
870 GeV.  Measurements of jet spectra and angular distributions from
runs IA and IB at the Tevatron, from future Tevatron runs, and
eventually from the LHC will shed further light on the model.

This theory would be even more interesting if it could also shed light
on the origins of electroweak and flavor symmetry breaking.  While the
minimal model described here has little connection to these issues, it appears
possible to include the extended strong interactions within a
framework that addresses them\cite{colph}.  Work on these questions is in
progress. 

\section*{Acknowledgments} 
We thank R.S. Chivukula and R.M. Harris for conversations; the Aspen
Center for Physics and the Fermilab Summer Visitors Program for hospitality; and
the NSF Faculty Early Career Development (CAREER) program, the DOE Outstanding
Junior Investigator program, and the JSPS Invitation Fellowship program for
financial support.  {\em This work was supported in part by the NSF under grant
PHY-95-1249 and by the DOE under grant DE-FG02-91ER40676.}



\section*{References} 
\begin{thebibliography}{99} 

\bibitem{WARSAW96} ``A Combination of Preliminary Electroweak
  Measurements and Limits on the Standard Model'', The LEP Experiments,
  the LEP Electroweak Working Group, and the SLC Heavy Flavour Group,
  CERN-PPE/96-183 (1996).

\bibitem{CDFexc} ``Inclusive Jet Cross Section in $\bar p p$
Collisions at $\sqrt{s} = 1.8$ TeV'', CDF Collaboration, F.~Abe {\it
et al.}, FERMILAB-PUB-96/020-E, hep-ex/9601008.

\bibitem{newint} R.S. Chivukula, A.G. Cohen, and E.H. Simmons,
Phys. Lett. B380 (1996) 92, hep-ph/9603311.

\bibitem{topglu} C.T. Hill Phys. Lett. {\bf B266} (1991) 419;
C.T. Hill and S.J. Parke, Phys. Rev. {\bf D49} (1994) 4454.

\bibitem{pdfs} ``Improved Parton Distributions from Global Analysis if
  Recent Deep Inelastic Scattering and Inclusive Jet Data,'' H.L. Lai et
  al., MSUHEP-60426, CTEQ-604,(1996), hep-ph/9606399; M. Klasen and G.
  Kramer, Phys. Lett. B386 (1996) 384, hep-ph/9605210; P. Chiapetta, J.
  Layssac, F.M. Renard, and C. Versegnassi, Phys. Rev. D54 (1996) 789,
  hep-ph/9601306; G. Altarelli, N. Di Bartolomeo, F. Feruglio, R. Gatto,
  and M.L. Mangano, Phys. Lett. B375 (1996) 292, hep-ph/9601324; M.
  Bander, Phys. Rev. Lett. 77 (1996) 601, hep-ph/9602330; ``High $E_T$
  Jets at $p\bar p$ Collisions and Triple Gauge Vertex'', B.A.  Arbuzov,
  hep-ph/9602416; Z. Bern, A.K. Grant, and A.G. Morgan, Phys. Lett. B387
  (1996) 804, hep-ph/9606466.

\bibitem{colph} ``Coloron Phenomenology'', E.H. Simmons, (1996).  BUHEP-96-24.
hep-ph/9608269, Phys. Rev. D (to appear); ``Limits on Flavor-Universal
Colorons'', E.H. Simmons, Proceedings of Snowmass 1996, hep-ph/9608349.

\bibitem{CDFdij} CDF Collaboration (F. Abe et al.)
Phys. Rev. Lett. {\bf 74} (1995) 3538.  hep-ex/9501001.

\bibitem{axig} J. Pati and A. Salam, Phys. Lett. {\bf 58B} (1975) 333;
J. Preskill, Nucl. Phys. {\bf B177} (1981) 21; L. Hall and A. Nelson,
Phys. Lett. {\bf 153B} (1985) 430; P.H. Frampton and S.L. Glashow,
Phys. Lett. {\bf B190} (1987) 157 and Phys. Rev. Lett. {\bf 58} (1987)
2168; J. Bagger, C. Schmidt, and S. King, Phys. Rev. 
{\bf D37} (1988) 1188. 

\bibitem{ua1} C. Albajar et al. (UA1 Collaboration), Phys. Lett. {\bf
B209} (1988) 127.

\bibitem{cdt} R.S. Chivukula, B.A. Dobrescu, and J. Terning, Phys.
Lett. {\bf B353} (1995) 289.

\bibitem{mrs_pak} A.D. Martin, R.G. Roberts and W.J. Stirling,
Phys. Lett. {\bf B306} (1993) 145.

\bibitem{rmh} R.M. Harris, private communication.  See preliminary CDF
results at http://www-cdf.fnal.gov/physics/new/qcd/qcd\_plots/two jet/public/dijet\_new\_physics.html\
\ \ on the World Wide Web.

\bibitem{tc2} C.T. Hill, Phys. Lett. {\bf B345} 483 (1995).


\end{thebibliography}
\end{document}